\documentclass[12pt,preprint]{aastex}

\shorttitle{Spectroscopy of SN 2002ic}

\shortauthors{Deng et al.}

\begin{document}

\title{Subaru Spectroscopy of the Interacting Type Ia Supernova SN
2002ic: Evidence of a Hydrogen-rich, Asymmetric Circumstellar
Medium}

\author{J.~Deng\altaffilmark{1,2},
        K.~S.~Kawabata\altaffilmark{3,4},
        Y.~Ohyama\altaffilmark{5},
        K.~Nomoto\altaffilmark{1,2},
        P.~A.~Mazzali\altaffilmark{1,2,6},
        L.~Wang\altaffilmark{7},
        D.~J.~Jeffery\altaffilmark{8},
        M.~Iye\altaffilmark{4,9},
        H.~Tomita\altaffilmark{2},
          and
        Y.~Yoshii\altaffilmark{1,10}
}

\altaffiltext{1}{Research Center for the Early Universe,
University of Tokyo, Hongo 7-3-1, Bunkyo-ku, Tokyo 113-0033,
Japan; deng@astron.s.u-tokyo.ac.jp.}

\altaffiltext{2}{Department of Astronomy, University of Tokyo,
Hongo 7-3-1, Bunkyo-ku, Tokyo 113-0033, Japan.}

\altaffiltext{3}{Department of Physical Sciences, Hiroshima
University, Hiroshima 739-8526, Japan.}

\altaffiltext{4}{Optical and Infrared Astronomy Division, National
Astronomical Observatory of Japan, 2-21-1 Ohsawa, Mitaka, Tokyo
181-8588, Japan.}

\altaffiltext{5}{Subaru Telescope, National Astronomical
Observatory of Japan, 650 North A'ohoku Place, Hilo, HI 96720.}

\altaffiltext{6}{INAF-Osservatorio Astronomico, Via Tiepolo, 11,
 34131 Trieste, Italy.}

\altaffiltext{7}{Lawrence Berkeley National Laboratory, 1
Cyclotron Road, Berkeley, CA 94720.}

\altaffiltext{8}{Department of Physics, University of Nevada at
Las Vegas, Box 454002, Las Vegas, NV 89154.}

\altaffiltext{9}{Department of Astronomy, Graduate University
 for Advanced Studies, Mitaka, Tokyo 181-8588, Japan.}

\altaffiltext{10}{Institute of Astronomy, University of Tokyo,
Mitaka, Tokyo 181-0015, Japan.}

\begin{abstract}

Optical spectroscopy of the Type Ia supernova SN 2002ic obtained
on 2003 June 27.6~UT, i.e., $\sim$ 222 rest-frame days after
explosion, is presented. Strong H emission indicates an
interaction between the expanding SN ejecta and an H-rich
circumstellar medium (CSM). The spectrum of SN~2002ic resembles
those of SNe 1997cy and 1999E. The three SNe also have similar
luminosities, suggesting that they are the same phenomenon and
that the CSM is also similar. We propose a new classification,
Type IIa SNe, for these events. The observed line profiles and
line ratios are measured and discussed within the ejecta-CSM
interaction scenario. The emission in H Balmer, [\ion{O}{3}], and
\ion{He}{1} lines, and in permitted \ion{Fe}{2} blends, resembles
the spectra of the Type IIn SN 1987F and of Seyfert 1 galaxies. A
high-density, clumpy CSM is inferred. Strong, very broad
[\ion{Ca}{2}]/\ion{Ca}{2} and [\ion{O}{1}]/\ion{O}{1} emissions
imply that not all the outer SN ejecta were decelerated in the
interaction, suggesting that the CSM is aspherical.

\end{abstract}


\section{INTRODUCTION}

\citet{ham03} reported strong \ion{Fe}{3}, \ion{Si}{2}, and
\ion{S}{2} features in the early-time spectra of SN~2002ic and
classified it as a Type Ia supernova (SN Ia). However, strong
H$\alpha$ emission was also observed. The detection of H$\alpha$
is unprecedented in an SN Ia. (For reviews on SN spectra, see
\citealt{fil97}). The emission was broad (FWHM $> 1000$
km~s$^{-1}$), suggesting that it was intrinsic to the SN.
\citet{ham03} suggested that it arose from the interaction between
the SN ejecta and a dense, H-rich circumstellar medium (CSM), as
in SNe IIn (e.g., \citealt{chu91,che94}). If this interpretation
is correct, SN~2002ic may be the first SN Ia to show direct
evidence of the circumstellar (CS) gas ejected by the progenitor
system, presenting a unique opportunity to explore the CSM around
an SN Ia and the nature of the progenitor system.

In this Letter, we present optical spectroscopy of SN~2002ic
obtained more than 200 days after explosion, and discuss it within
the context of the ejecta-CSM interaction.

\section{OBSERVATIONS AND RESULTS}

Observations were carried out on 2003 June 27.6~UT with the Faint
Object Camera and Spectrograph (\citealt{kas02}) attached to the
Cassegrain focus of the 8.2 m Subaru Telescope. For the red
observation (5900 -- 10200 \AA), we used a 300 groove mm$^{-1}$
grism (centered at 7500 \AA) and an order-cut filter O58. For the
blue observation (3800 -- 7000 \AA), we used another 300 grooves
mm$^{-1}$ grism (centered at 5500 \AA) and no filter. A $0\farcs
8$ width slit was used under moderate seeing conditions (FWHM
$\simeq 0\farcs 6-0\farcs 7$), resulting in a spectral resolution
$\lambda/\Delta\lambda \sim 650$ ($\sim 460$ km~s$^{-1}$). The
total exposure time was 1680 s for each observation. The flux was
calibrated using observations of BD+28$\arcdeg$4211 \citep{mas90}.
A systematic error of $\sim$ 0.1 mag is caused by the insufficient
width of the slit compared with the seeing size. Although all data
were taken in polarimetric mode, the signal-to-noise ratio is not
high enough for sufficiently accurate polarimetry ($\lesssim 0.2$
\%). Therefore, in the following only the flux data are discussed.

The spectrum is shown in Figure \ref{fig1}, after correction for
redshift ($z=0.0666$; \citealt{ham03}) and Galactic extinction
($E_{B-V}=0.06$; \citealt{sch98}). The epoch is $\sim 222$
rest-frame days after explosion, which we assume occurred on 2002
November 3 UT \citep{ham03}. The spectrum has a brightness of
$m_V\sim 19.0$. In other words, the SN is only $\sim 1.5$ mag
fainter than at maximum light \citep{ham03}, while a normal SN Ia
at similar epochs would have faded by $\sim$ 5 mag with respect to
maximum light (Figure \ref{fig2}).


The spectrum is strikingly similar to those of the peculiar
SNe~1997cy \citep{tur00} and 1999E \citep{rig03}, which were
classified as Type IIn. (See also \citealt{wan04}.) SN~2002ic
would also have been so classified, had it not been discovered at
an early epoch when it appeared to be a genuine SN Ia.
\cite{ham03} also noticed similarities to SN~1997cy in an earlier
spectrum ($\sim$ 71 days after explosion).

The $UBVRI$ light curves (LCs) of the three SNe are also similar
(Figure \ref{fig2}). To construct the LC of SN~2002ic, we first
integrated the Subaru spectrum. This yielded $L = (5.8\pm
0.6)\times 10^{42}$ ergs~s$^{-1}$, assuming a distance of 307 Mpc.
The bolometric correction thus estimated was used to convert $m_V$
at earlier phases in \citet{ham03} and the late-time MAGNUM
telescope \citep{yos02} photometry into rough bolometric
luminosities.


\section{SPECTRAL ANALYSIS}

The H$\alpha$ profile was decomposed using three Gaussians (Figure
\ref{fig3}, {\em top left}). (We used Gaussians for mathematical
convenience, although real component profiles can be different.) A
narrow core (FWHM $\sim 1000$ km~s$^{-1}$) is unresolved owing to
low instrumental resolution. The intermediate component has FWHM
$\sim 4800$ km~s$^{-1}$. It may develop from the $\sim 1800$
km~s$^{-1}$ component seen at earlier phases \citep{ham03}. The
broad blue wing centered at $\sim 6350$~\AA\ is likely
[\ion{O}{1}] $\lambda\lambda$6300,6364 (FWHM $\sim 26000$
km~s$^{-1}$). The integrated fluxes are $4\times 10^{-15}$,
$1.5\times 10^{-14}$, and $3.2\times 10^{-14}$
ergs~s$^{-1}$~cm$^{-2}$, respectively.

Broad [\ion{O}{1}] $\lambda\lambda$6300, 6364 is also present in
SNe~1997cy and 1999E. [\ion{O}{1}] lines seen in a few SNe IIn
(e.g., \citealt{fra02}) are narrower than 4000 km~s$^{-1}$ and
much weaker. [\ion{O}{1}] $\lambda$5577 is not obvious. Its
potential location, marked by an arrow in Figure \ref{fig1}, is
well in the smoothly declining part of the $\sim$ 5100 -- 5600
\AA\ feature.

The H$\beta$ $\lambda$4861 -- [\ion{O}{3}] $\lambda$5007 complex
was decomposed with four Gaussians (Figure \ref{fig3}, {\em top
right}). The two H$\beta$ components have FWHM $\sim$ 1700 and
4000 km~s$^{-1}$ and flux of $\sim 8\times 10^{-16}$ and $9\times
10^{-16}$ ergs~s$^{-1}$~cm$^{-2}$, respectively. The unresolved
[\ion{O}{3}] $\lambda$5007 component (FWHM $\sim 500$ km~s$^{-1}$)
has a flux of $\sim 1.4\times 10^{-16}$ ergs~s$^{-1}$~cm$^{-2}$.
The expected intensity of [\ion{O}{3}] $\lambda$4959 is only
one-third that of $\lambda$5007 \citep{ost89}. So the $\sim$4950
\AA\ feature (FWHM $\sim 6000$ km~s$^{-1}$) is mainly due to
\ion{Fe}{2} multiplet 42.

Our spectrum shows a possible [\ion{O}{3}] $\lambda$4363 line
(Figure \ref{fig3}, {\em bottom right}), as in SNe~1997cy and
1999E. A tentative Gaussian fit gives FWHM $\sim 1200$ km~s$^{-1}$
and a flux of $\sim 1.7\times 10^{-16}$ ergs~s$^{-1}$~cm$^{-2}$,
which are likely overestimates. The feature to the left is also
unresolved: it seems not to be H$\gamma$ $\lambda$4340; the
SN~1999E spectrum ({\em open circles}) does not show H$\gamma$.

We decomposed the 7300 \AA\ feature into \ion{He}{1}
$\lambda$7065, [\ion{Ca}{2}] $\lambda\lambda$7291,
7324/[\ion{O}{2}] $\lambda\lambda$7320, 7330, and \ion{O}{1}
$\lambda$7774 (Figure \ref{fig3}, {\em bottom left}). The FWHMs
are $\sim$ 3000, 18,000, and 10,000 km~s$^{-1}$, and the fluxes
are $\sim 6\times 10^{-16}$, $1.2\times 10^{-14}$, and $2.3\times
10^{-15}$ ergs~s$^{-1}$~cm$^{-2}$, respectively. \ion{He}{1}
$\lambda$5876 is weak (FWHM $\sim 1400$ km~s$^{-1}$, flux $\sim
2\times 10^{-16}$ ergs~s$^{-1}$~cm$^{-2}$), as in SN 1997cy.

The strong emission near 8500~\AA\ is a blend of the \ion{Ca}{2}
IR triplet and \ion{O}{1} $\lambda$8446 (FWHM $\sim 13000$
km~s$^{-1}$ and flux $\sim 4.7\times 10^{-14}$
ergs~s$^{-1}$~cm$^{-2}$). A broad feature to the red could be weak
\ion{O}{1} $\lambda$9265. The \ion{Ca}{2} H and K flux is $\sim
1\times 10^{-14}$ ergs~s$^{-1}$~cm$^{-2}$.


The lines we have identified can be divided into two groups based
on their width. One group, comprising the H Balmer, [\ion{O}{3}],
and \ion{He}{1} lines, have unresolved cores (FWHM $\lesssim 1000$
km~s$^{-1}$) and, in the case of H$\alpha$ and H$\beta$,
components of intermediate width (FWHM $\sim 3000 - 5000$
km~s$^{-1}$). The other group includes broad
[\ion{Ca}{2}]/\ion{Ca}{2} and [\ion{O}{1}]/\ion{O}{1} lines (FWHM
$>10000$ km~s$^{-1}$). These lines do not show a narrow component.

We believe the unresolved components are CSM emissions, likely
from a progenitor wind, as in SNe IIn (e.g., \citealt{fra02}). R.
Kotak \& W. P. S. Meikle (2004, in preparation), using
high-resolution spectroscopy, saw a P Cygni line with absorption
velocity $\sim 100$ km~s$^{-1}$ atop the H$\alpha$ core.
\citet{rig03} measured a similar P Cygni velocity for H$\alpha$
($\sim 200$ km~s$^{-1}$) in SN~1999E. The intermediate components
of H$\alpha$ and H$\beta$ may be formed by multiple Thomson
scattering of narrow emissions in CSM clouds of $n\gtrsim 10^8$
cm$^{-3}$ \citep{wan04}. The [\ion{O}{3}]-emitting region may have
$n\sim 10^7$ cm$^{-3}$ ($T\sim 10^4$ K), near the critical density
of the [\ion{O}{3}] $2p^{2}~^{1}S$ level, as implied by the
comparable flux of the possible [\ion{O}{3}] $\lambda$4363 line
\citep{ost89}.

The total H$\alpha$ luminosity, $L({\rm H_\alpha})\sim 2-3\times
10^{41}$ ergs~s$^{-1}$, may imply $\sim 0.4-3 M_\sun$ of
high-density ionized H ($\sim 10^8-10^9$ cm$^{-3}$) in the CSM,
assuming case B recombination (see also \citealt{wan04}). Our
Balmer decrement [$L({\rm H_\alpha})/L({\rm H_\beta})\sim 10$] is
much steeper than expected for case B. Either collisional
processes are important or other Balmer photons are absorbed and
cascaded into H$\alpha$, or both. In either case, a high density
is implied.

We suggest that the broad O/Ca lines are emitted by the SN Ia
ejecta. In the W7 model for SNe Ia \citep{nom84}, the outer C/O
layer has $v > 12000$ km~s$^{-1}$ and a typical density of $\sim
10^5$ cm$^{-3}$ around 200 days, assuming free expansion. This is
consistent with the detection of [\ion{O}{1}]
$\lambda\lambda$6300, 6364 but not of [\ion{O}{1}] $\lambda$5577,
which suggests $n<10^8/t$(days) cm$^{-3}$ if $T\sim 10^4$ K (see
Figure 7 in \citealt{lei91}). The weakness of the [\ion{Ca}{2}]
$\sim$ 7300~\AA\ line with respect to the IR triplet may suggest
that the density is not as low. However, the flux ratio
$F_{7300}/F_{8500}$ in SN~1997cy actually decreases with time and
density. Perhaps the $\sim$ 8500~\AA\ feature is dominated by
Ly$\beta$-pumped \ion{O}{1} $\lambda$8446 at later phases.

Parts of the outer ejecta must avoid strong CS interaction to
retain high velocities and produce the broad O and Ca emissions (
powered by X-ray/UV radiation from the interaction region). The
velocity of the shocked region and the preshocked ejecta is too
low to explain the width of these lines. According to
hydrodynamical simulations (T. Suzuki et~al. 2004, in
preparation), the pre-shocked ejecta have velocities less than
7000 km~s$^{-1}$ at $\sim 100$ days after explosion and less than
4000 km~s$^{-1}$ at $\sim 200$ days, and the shocked ejecta are
decelerated to similar velocities. We suggest that the CSM is
concentrated near the equator, so that the ejecta near the pole do
not strongly interact with it (see also Section~\ref{sec-disc}).

We suggest that the features near 4300, 4600, 4950, and 5300 \AA\
(marked by circles and one arrow in Figure \ref{fig4}) are broad
blends of \ion{Fe}{2} multiplets 27; 38 and 37; 42; and 49, 48,
and 42, respectively. They strikingly resemble the \ion{Fe}{2}
permitted emissions in Seyfert 1 galaxies (\citealt{ost89}; see
also \citealt{fil89} for similar conclusions in SN IIn 1987F).
They may come from the cool shell in the reverse shocked region
or, more likely, from the dense CSM clouds, suggesting a density
$\gtrsim 10^9$ cm$^{-3}$.

\section{SPECTRAL COMPARISON WITH OTHER TYPES} \label{sec-comp}


In Figure \ref{fig4}, we compare the late-time spectra of
SNe~2002ic, 2000cx (Ia), 1998bw (Ic), and 1988Z (IIn). Strong
[\ion{Fe}{3}]/[\ion{Fe}{2}] blends dominate the normal SNe Ia
spectra (e.g., \citealt{liu97}) and this suggests low density and
high ionization relative to SN 2002ic where we find permitted Fe
II lines. SNe Ic  show very strong \ion{Mg}{1}] $\lambda$4571
\citep{maz01}. This line could blend with the \ion{Fe}{2} feature
near 4400 -- 4700 \AA\ in SN~2002ic, but SNe 1997cy and 1999E
disfavor a strong \ion{Mg}{1}] contribution since in their spectra
the central wavelength is $\sim$ 4640 \AA.

The spectrum of the SN IIn 1988Z also looks different from that of
SN~2002ic, although both show strong H$\alpha$ emission. In
SN~1988Z, broad [\ion{O}{1}] $\lambda\lambda$6300, 6364 and
[\ion{Ca}{2}] $\lambda\lambda$7291, 7324 are absent, as in other
SNe IIn \citep{fil97}, while \ion{Ca}{2} IR/\ion{O}{1} emission is
weak. The  broad Fe features near 4600 and 5300 \AA\ are also
absent in SN~1988Z.

\section{DISCUSSION} \label{sec-disc}

The currently preferred model for SNe Ia is the thermonuclear
explosion of a C+O white dwarf (WD) in a binary system, reaching
the Chandrasekhar limit via either accretion from a normal
companion (the SD scenario, which is generally favored) or merging
with another WD (the DD scenario). (For recent reviews, see
\citealt{nom00,liv00}.)

The SD scenario predicts the presence of an H/He-rich CSM. The
discovery of strong CSM interaction in SN~2002ic may prove that
this scenario does exist in nature. SNe~1997cy and 1999E may also
be CS-interacting SNe Ia. However, such events are rare. Solid
observational evidence of CSM has not yet been found in other SNe
Ia \citep{lun03}.

Our spectral analysis suggests a high-density \ion{H}{1}-emitting
CSM ($n \sim 10^8-10^9$~cm$^{-3}$); so this is probably clumpy.
Assuming the CSM was formed in a progenitor wind, we relate the
mass-loss rate $\dot{M}$ and the wind velocity $u$ to the
H$\alpha$ luminosity, through $L({\rm H_\alpha})\sim 1-5\times
10^{39}$ $(\dot{M}/10^{-2}{\rm~M_\sun~yr^{-1}})^2$ $(u/100
{\rm~km~s^{-1}})^{-2}$ $r_{16}^{-1} f^{-1}$ ergs~s$^{-1}$, where
$r_{16}$ is the radius of the CSM shock in units of $10^{16}$ cm
and $f$ is the CSM filling factor. For $u\sim 100$ km s$^{-1}$ and
$f\sim 0.01$, $\dot{M}$ can be as high as $\sim 10^{-2}$
M$_\sun$~yr$^{-1}$. The case B emissivity used here may
underestimate $L({\rm H_\alpha})$ by $\sim 50\%$, considering that
most H$\beta$ photons may have cascaded to H$\alpha$.

We have identified high-velocity lines (FWHM $\gtrsim 10^4$
km$^{-1}$) emitted in the ejecta of SN~2002ic. Based on the
hydrodynamical model of T. Suzuki et~al. (2004, in preparation),
we suggest an equatorially concentrated structure for the CSM to
explain the coexistence of these lines with the strong CS
interaction. Such a geometry is not unexpected for the mass loss
from a binary system or for stars approaching the end of the
asymptotic giant branch (AGB). A preexisting clumpy disk was also
suggested by \citet{wan04}, based on spectropolarimetry.

Is $\dot{M}\sim 10^{-2}$ M$_\sun$~yr$^{-1}$ too high for the SD
scenario? The highest $\dot{M}$ observed in AGB stars is between
$10^{-3}$ and $10^{-4}$ M$_\sun$~yr$^{-1}$ \citep{ibe95};
$\dot{M}$ in our estimate is scaled as $u$. The bulk of the CSM
may be near the equator and at low velocity ($\sim 10$
km~s$^{-1}$), which cannot be resolved. The observed H$\alpha$ P
Cygni, dominated by the emission component, could be produced by
some high-velocity CSM ($\gtrsim 100$ km~s$^{-1}$) along the pole,
which should be close to our line of sight. Similar wind patterns
have been observed in some symbiotic stars (e.g.,
\citealt{sol85}), which have been suggested as candidates for SN
Ia progenitors.

Further observations and modeling are required to understand the
nature of the CSM giving rise to this type of event, which may be
classified as ``Type IIa.'' Suggestions include a common envelope
\citep{liv03}, WD accretion wind \citep{hac99}, and the superwind
from an AGB star exploding
 as a ``type 1.5'' event \citep{ham03}.

\acknowledgments

We thank the Subaru Telescope staff for their kind support, R.
Kotak and P. Meikle for allowing us to cite their results, and M.
Turatto for data of SNe~1997cy and 1999E.


\clearpage

\figcaption[f1.eps]{Spectra of SNe~2002ic ({\em thick lines};
$\sim 222$ d), 1997cy ({\em thin lines}; \citealt{tur00}), and
1999E ({\em dashed line}; \citealt{rig03}). \label{fig1}}

\figcaption[f2.eps]{Comparison of the $UBVRI$ LC of SN~2002ic with
those of SNe~1997cy \citep{tur00}, 1999E \citep{rig03}, and the
normal SN Ia 1994D \citep{con00}. \label{fig2}}

\figcaption[f3.eps]{Gaussian decomposition of line profiles. Solid
squares show the observed profiles, thin lines are the Gaussian
components, and thick lines are the combined Gaussian profiles.
Open circles show [\ion{O}{3}] $\lambda$4363 in SN~1999E.
\label{fig3}}

\figcaption[f4.eps]{Spectra of SN~2002ic ({\em thick lines}; $\sim
222$ d) and SNe Ia 2000cx ({\em top}; \citealt{li01}), Ic 1998bw
({\em middle}; \citealt{pat01}), and IIn 1988Z ({\em bottom};
\citealt{tur93}). \label{fig4}}

\clearpage

\epsscale{0.8}
\plotone{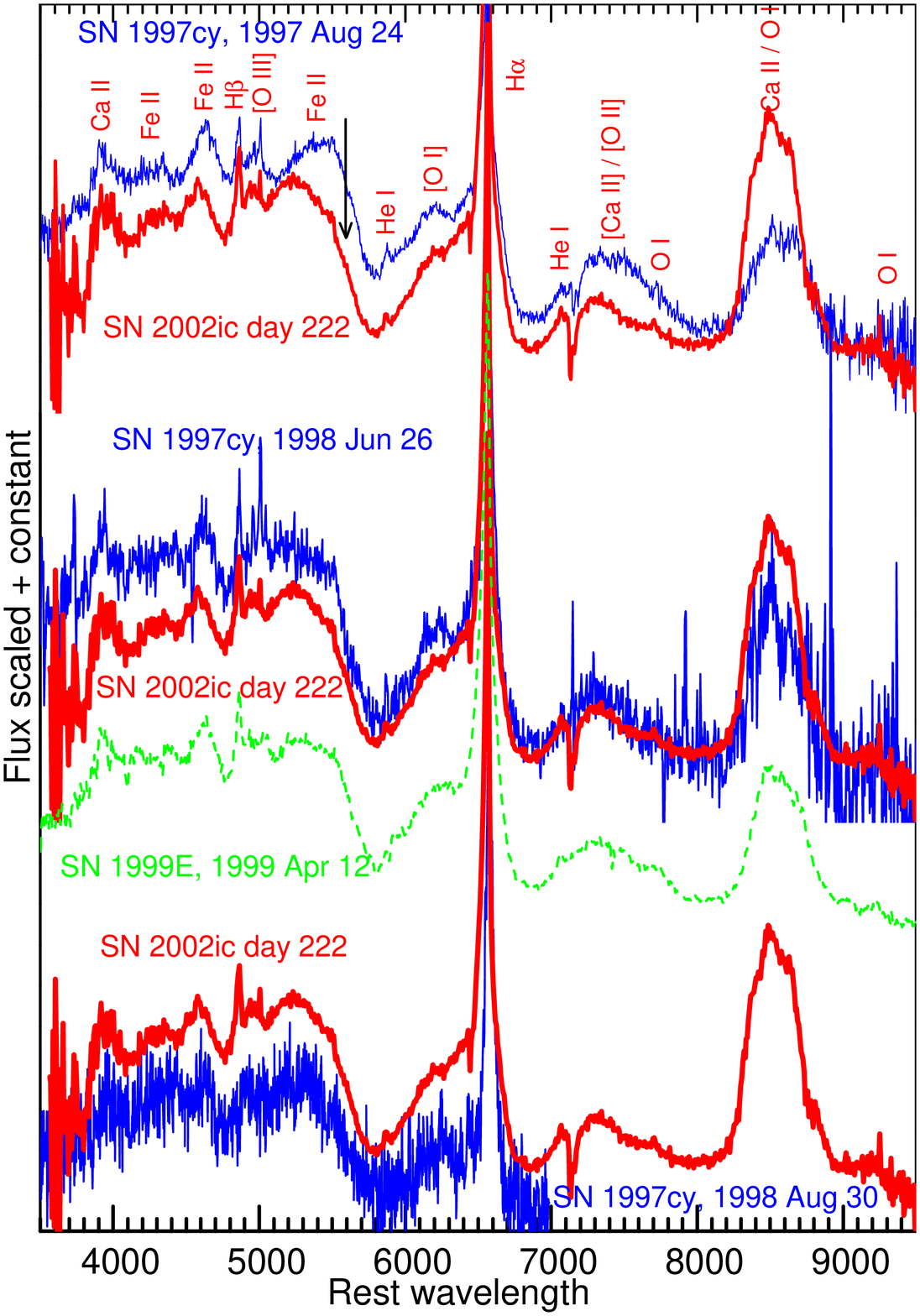} \clearpage

\plotone{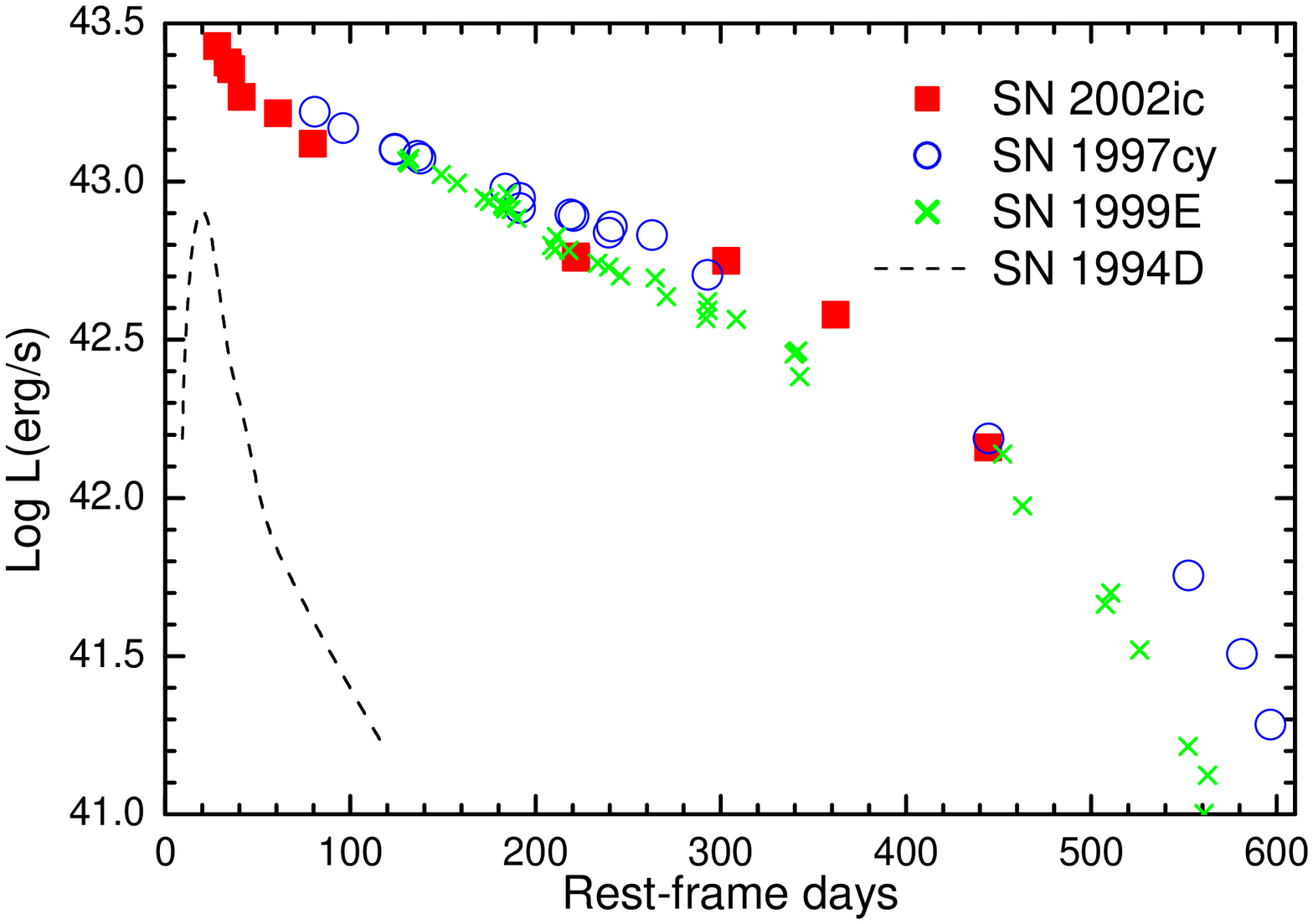} \clearpage

\plotone{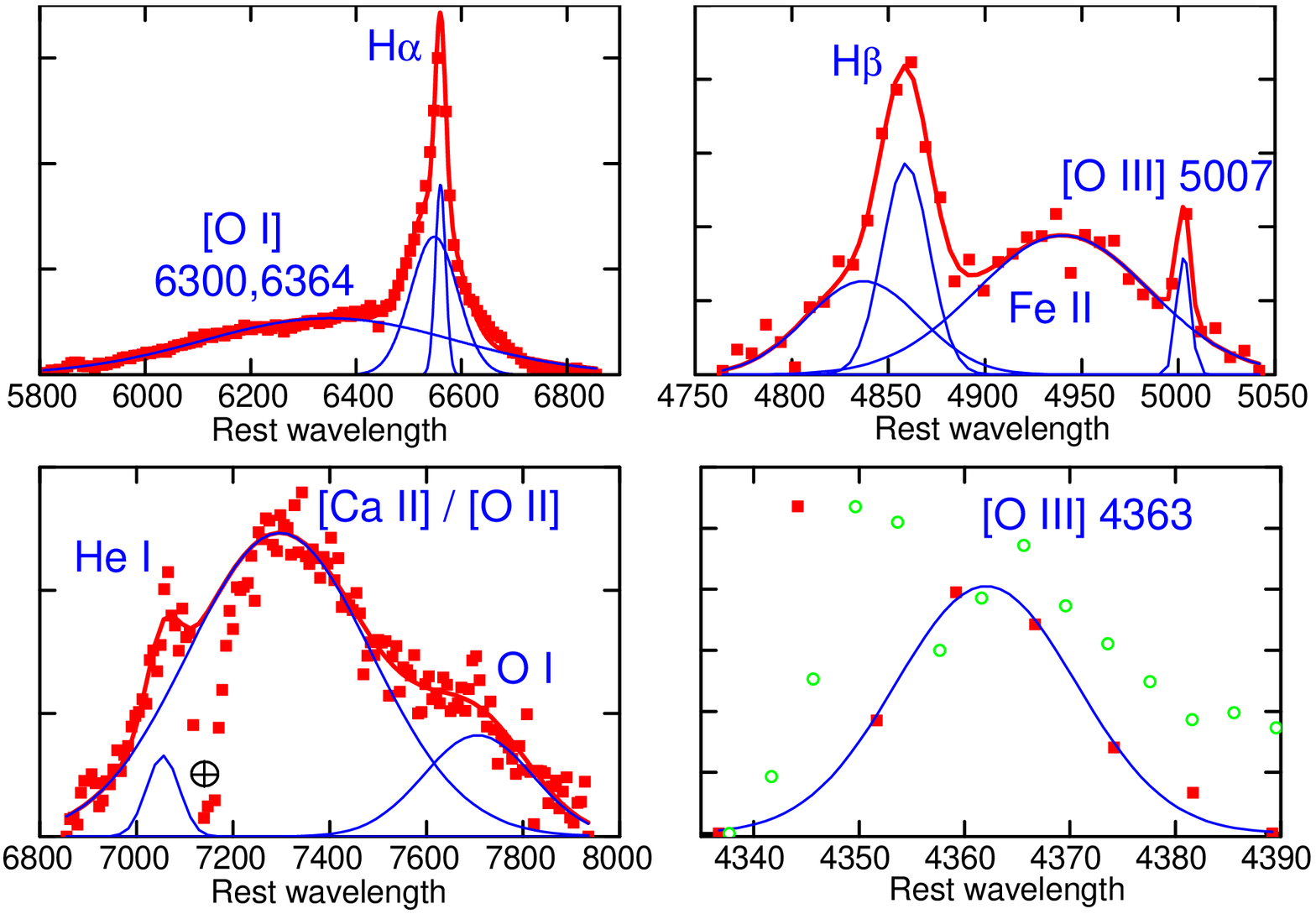}

\plotone{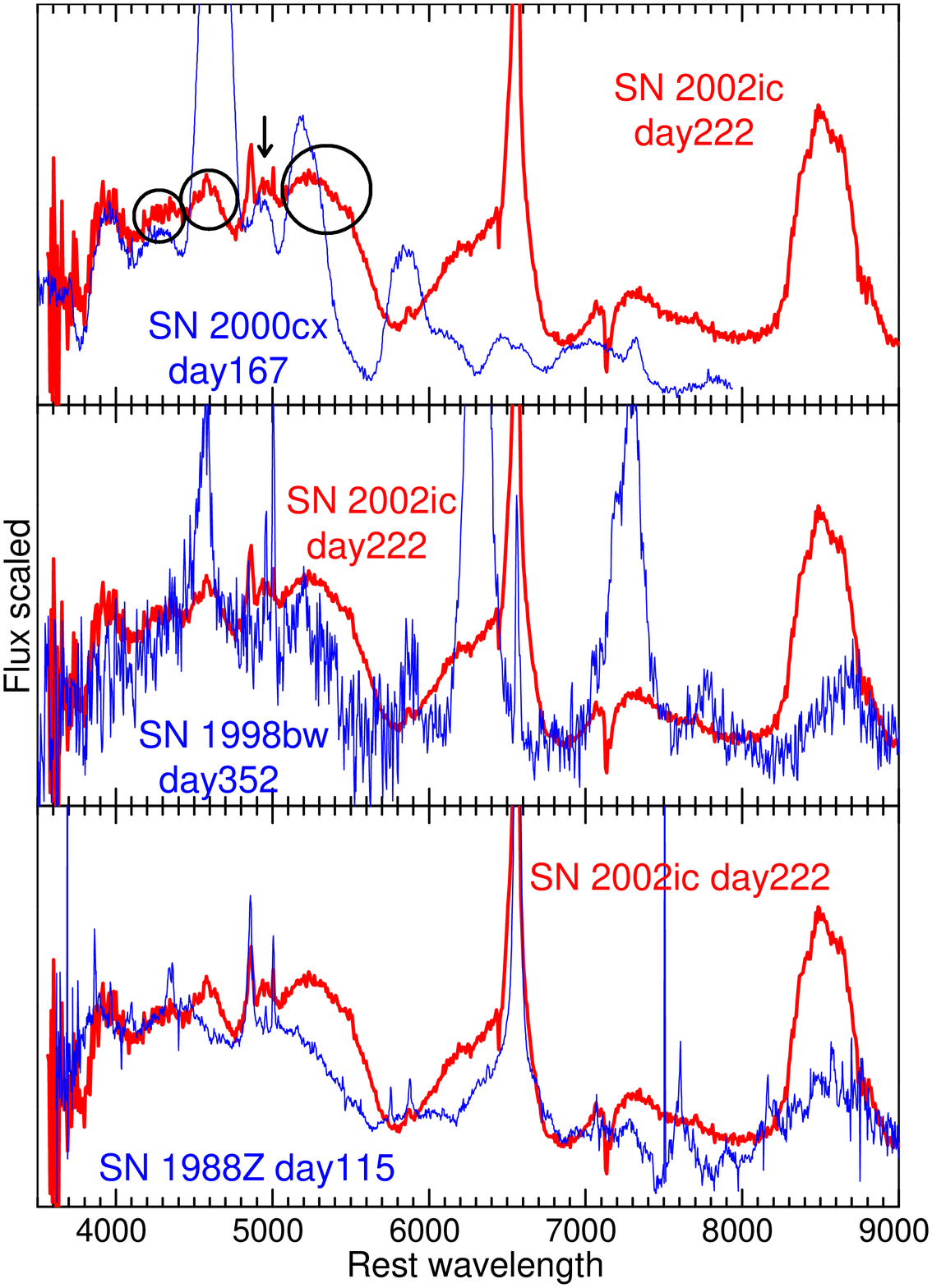}

\end{document}